\documentclass[a4paper]{article}
\usepackage{textpos}
\usepackage{INTERSPEECH2020}
\usepackage{multirow}
\usepackage{enumitem}
\usepackage{svg}
\usepackage{hyperref}
\usepackage{subcaption}

\newlist{steps}{enumerate}{1}
\setlist[steps, 1]{label = Step \arabic*:}
\usepackage[ruled,vlined]{algorithm2e}
\usepackage[compress]{cite}
\usepackage{amsmath}

\title{Improve few-shot voice cloning using multi-modal learning}
\name{Haitong Zhang, Yue Lin}
\address{
  NetEase Games AI Lab, China \\
  \{zhanghaitong01, gzlinyue\}@corp.netease.com}

\begin{document}

\maketitle

\begin{abstract}
Recently, few-shot voice cloning has achieved a significant improvement. However, most models for few-shot voice cloning are single-modal, and multi-modal few-shot voice cloning has been understudied. In this paper, we propose to use multi-modal learning to improve the few-shot voice cloning performance. Inspired by the recent works on unsupervised speech representation, the proposed multi-modal system is built by extending Tacotron2 with an unsupervised speech representation module. We evaluate our proposed system in two few-shot voice cloning scenarios, namely few-shot text-to-speech~(TTS) and voice conversion~(VC). Experimental results demonstrate that the proposed multi-modal learning can significantly improve the few-shot voice cloning performance over their counterpart single-modal systems.


\end{abstract}
\noindent\textbf{Index Terms}: few-shot voice cloning, text-to-speech, voice conversion, multi-modal

\section{Introduction}



Recently, text-to-speech has witnessed a great improvement due to the development of the sequence-to-sequence models \cite{wang2017tacotron, shen2018natural} and high-fidelity neural vocoders \cite{vanwavenet, oord2018parallel}. Meanwhile, voice conversion has gained rapid development using various techniques \cite{zhao2020voice}.

With the rapid development of TTS and VC, few-shot voice cloning, namely cloning new voices with a small amount of data, has become an active research topic. For few-shot TTS, previous works have shown that fine-tuning either parts of the multi-speaker initial model or the whole model can provide high-quality results \cite{arik2018neural, chen2018sample, moss2020boffin}. Another approach is to train a speaker-adaptive model conditioned on a speaker embedding extracted from a pre-trained speaker recognition model \cite{ jia2018transfer, nachmani2018fitting}. This approach is useful when a few seconds of data is available and requires no fine-tune process. However, this approach has a drawback that the speaker similarity stops improving as more data is available \cite{chen2018sample}. Meanwhile, there are some works on building a voice conversion system for target speakers using a limited amount of data \cite{zhao2020voice}.

Although few-shot voice cloning has achieved a significant improvement, most systems for few-shot voice cloning are single-modal, and multi-modal few-shot voice cloning has been understudied. There are only a few previous works on multi-modal voice cloning. \cite{kim2020emotional, zhang2019joint, luong2020nautilus} proposed to receive multi-modal inputs using separate encoders so that the system can tackle TTS and VC simultaneously. A random masker or XOR training operator or the KL divergence between two encoder outputs is used to train the model for two tasks simultaneously. However, as found in \cite{kim2020emotional, zhang2019joint}, the TTS performance deteriorates while the VC results get better, indicating multi-modal learning for few-shot voice cloning is challenging while it may be helpful.


Inspired by previous works on unsupervised speech representation \cite{liu2020towards, chorowski2019unsupervised, zhang2020unsupervised}, this paper extends Tacotron2 (a state-of-the-art TTS system) with an unsupervised speech representation module to achieve multi-modal learning. To our best knowledge, this is the first work to leverage unsupervised speech representation to achieve multi-modal voice cloning. In addition, experimental results demonstrate that the proposed multi-modal system can significantly improve the few-shot voice cloning performance over its counterpart single-modal systems.

  \begin{figure*}[t]
    \centering
    \includegraphics[width=0.8\textwidth]{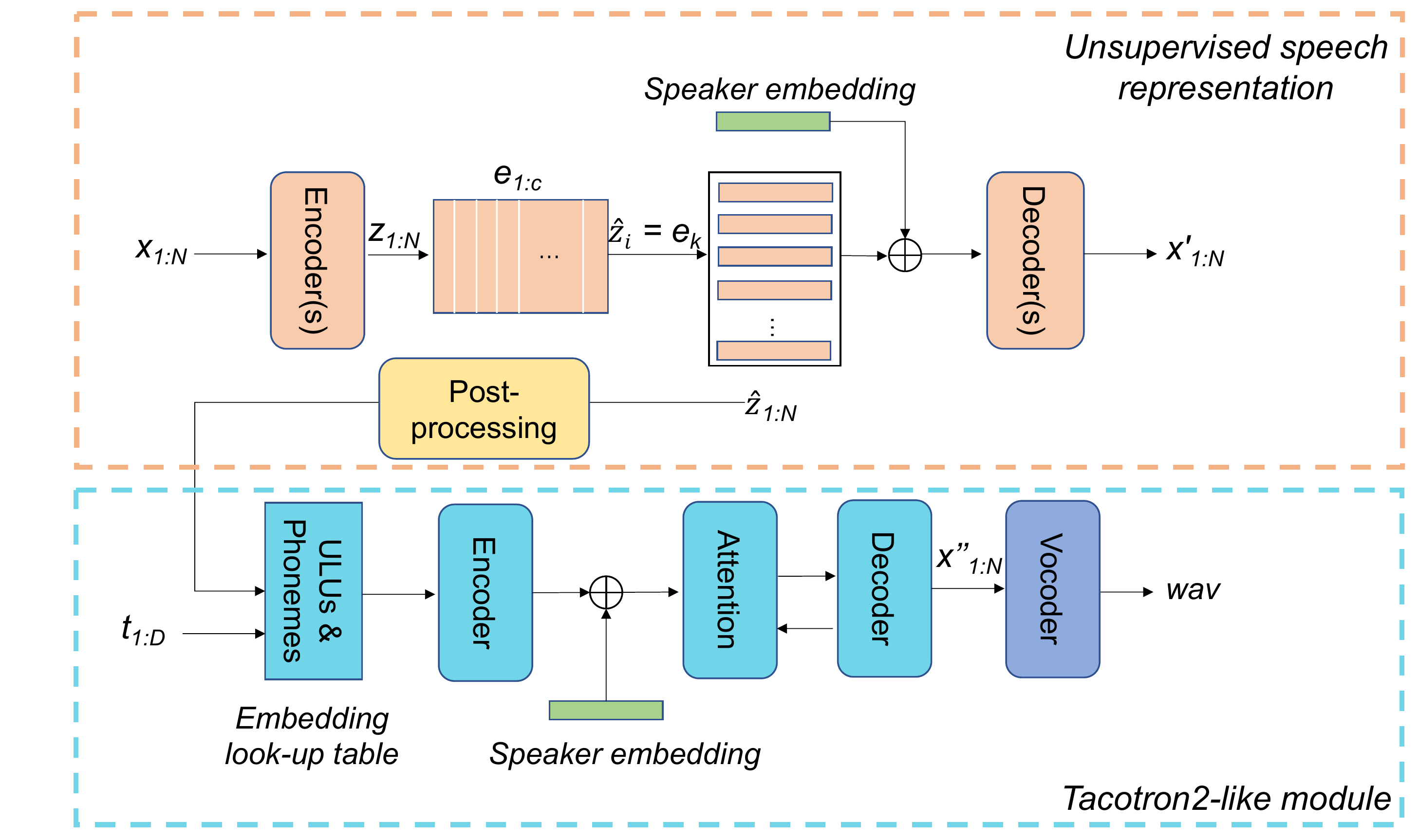}
    \caption{The proposed model structure.}
    \label{fig:model}
  \end{figure*}

\section{The Proposed System} \label{proposed}

\subsection{Motivation}
As found in \cite{liu2020towards}, the relationship between the unsupervised quantized speech representations matches the relationship between phonemes, indicating that the unsupervised quantized speech representations are similar to phonemes. Similarly, \cite{chorowski2019unsupervised} found that the quantized speech representations learned by VQ-VAE are consistently associated with phonemes. Thus, the quantized speech representations have been used in text-to-speech, voice conversion, and automatic speech recognition~\cite{liu2020towards, chorowski2019unsupervised, zhang2020neteasegames, zhang2020unsupervised}. Inspired by the previous works, this paper aims to improve few-shot voice cloning by extending the single-modal TTS model Tacotron2 with an unsupervised quantized speech representation module. 
 
The proposed system mainly consists of two modules, which is illustrated as Figure \ref{fig:model}. The first module is the unsupervised quantized speech representation module, which is used to extract unsupervised linguistic units given speech. The second module includes a Tacotron2-like module to generate mel-spectrogram conditioned on either unsupervised linguistic units or phonemes and a neural vocoder to re-construct waveform based on predicted mel-spectrogram. We call the proposed multi-modal system  MD-Tacotron in this paper.

\subsection{Unsupervised speech representation Module}

In this paper, we use VQ-VAE model \cite{van2017neural} as the extractor of discretized linguistic units due to its promising performance~\cite{van2017neural}. VQ-VAE is an auto-encoder model with a codebook dictionary  $e = C*D$ , where $C$ refers to the number of embeddings, and $D$ is the size of each embedding. The model also contains an encoder and decoder. The encoder takes speech features $x_{1:N}$ as inputs, and produces higher-level hidden representation $z_{1:N}$ before mapping into discretized embedding $\hat{z}_{1:N}$ in the dictionary by finding the nearest one as  $ \hat{z}_i = e_k$ , where $k = argmin_j$  $|| z_i - e_j||$, and $j \in  {1,2, \cdots ,C}$. Finally, the discretized embedding $\hat{z}_{1:N}$ and the speaker embedding $s$ are concatenated and passed into the decoder to reconstruct speech feature $x^{'}_{1:N}$. We use straight-through gradient estimation to approximate the gradients from the argmin operation \cite{van2017neural}. The loss function of the model is 

\begin{equation}
\begin{aligned}
L     &  \;  =  \;  ||  x_{1:N} \;  -  \; x^{'}_{1:N}  ||_2^2 \\
      &  \;  +  \;  ||  sg(z(x)) \;  -  \; e_j) ||_2^2  \\
      &  \;  +  \;  \beta \; * \; ||  z(x)  \; - \; sg(e_j) ||_2^2 
\end{aligned}
\end{equation}

where the first term is L2 loss for reconstructing the speech feature. The second term updates the codebook dictionary, with $sg$ denotes stop-gradient operation. The third term, as the commitment loss, encourages the encoder output $z$ to get close to the codebook embeddings, and the hyper-parameter $\beta$ balances the loss terms.

\subsection{Sequence-to-sequence module}

The second module is a seq-to-seq model to generate melspectrogram, and a neural vocoder to reconstruct the waveform conditioned on the predicted mel-spectrogram. We use Tacotron2 as our seq-to-seq module framework due to its outstanding performance in TTS \cite{shen2018natural}. To support multi-speaker training, we concatenated the encoder output with the speaker embedding extracted from the speaker embedding loop-up table. We also replaced the original attention with GMM attention for robust sequence modeling. In original Tacotron2, the model usually receives text representation as inputs, such as phonemes. In the proposed system, the Tacotron2-like module takes either unsupervised linguistic units (extracted from the VQ-VAE module) or phonemes as inputs. Unsupervised linguistic units (ULUs) and phonemes are embedded using an embedding lookup table. Specifically, the embedding lookup table is a  $(N_{ULUs} + N_{Phone} )  * N_{Dim}$  matrix, where $N_{ULUs}$, $N_{Phone}$, and $N_{Dim}$ refer to the number of unsupervised linguistic units, phonemes, and the embedding dimension, respectively. The module architecture is illustrated in the bottom part of Figure \ref{fig:model}.

We use Parallel WaveNet to generate waveform conditioned on mel-spectrogram. Inspired by \cite{ping2018clarinet}, we use a single Gaussian as the output distribution. We train a universal neural vocoder for all experiments in this paper.

\subsection{Training and inference}

Although the VQ-VAE and seq-to-seq module can be trained in an end-to-end mode, we use a two-stage training mode in this paper and left joint training for future investigation. In the first stage, we train the VQ-VAE module using the training data and use the trained VQ-VAE module to extract unsupervised linguistic units for the whole training data. We then apply the post-processing method to remove the consecutively repetitive  unsupervised linguistic units as in \cite{zhang2020unsupervised} to get a $<$unsupervised linguistic units (ULUs), audio$>$ paired data. We combine this dataset with the original $<$phonemes, audio$>$ paired data into the final dataset. In the second stage, we train the seq-to-seq module using this final dataset. At each training step, a batch of $<$ULUs, audio$>$ and $<$phonemes, audio$>$ is randomly sampled from the final dataset.

During inference, for VC, we first extract the unsupervised linguistic units for the input utterance using trained VQ-VAE module, and apply the post-processing method, then take the unsupervised units as the inputs of the Tacotron2-like module to generate waveform; for TTS, we use phonemes as the inputs of the seq-to-seq module to synthesize waveform.


\begin{table}
\caption{Results of objective and subjective evaluations for the few-shot TTS scenario.}
\begin{tabular}{  c c  c |  c c }
\hline 
\hline

  Model       & MCD   & WER    & NAT   &  SIM   \\ [0.5ex]
\hline 

GT        &   -      & 6.17   & $4.30 \pm .06$    &  $4.09 \pm .09$ \\ [0.3ex]
\hline


Tacotron  & 6.41   &  8.03   &  $3.80 \pm .07$   &  \textbf{3.66 $\pm$ .09} \\ 
 MD-Tacotron & \textbf{5.95}   & \textbf{6.93}   &  \textbf{3.90 $\pm$  .07 }    & $3.64 \pm .08$  \\ [0.1ex]
\hline


\end{tabular}
\label{table:SA}
\end{table}

\section{Experimental setup}

\subsection{Datasets and Training setup}

We evaluate the proposed method on voice cloning (including TTS and VC) using only 2-minute data from the target speakers. We implement experiments using VCTK data \cite{https://doi.org/10.7488/ds/1994}, which contains recordings from 108 speakers. We re-sample all waveforms into 16kHz.


Following \cite{zhang2020unsupervised}, we set the number of embedding $C$ and the size of each embedding $D$ in the VQ-VAE module into $256$ and $64$, respectively. The encoder contains 3 convolution layers with a kernel size of 5 and channel size of 512, and 2 BLSTM layers with a hidden size of 256. The decoder includes 3 BLSTM layers with a hidden size is 256. We used the Adam algorithm~for optimization. The initial learning rate is 1e-3, and we started decaying the learning rate from the 10k step with the decay step and rate is 15k and 0.5, respectively. We set $\beta$ into 0.25 as in \cite{van2017neural}. We set the output-per-step in the seq-to-seq module into $1$. We strongly encourage readers to refer to  \cite{shen2018natural} for other training setups.  


\subsection{Evaluation metrics}

We carry out both objective and subjective evaluations. For the objective evaluations, we compute the dynamic-time-warping (DTW) mel cepstrum distortion (MCD) \cite{kubichek1993mel}, a commonly used measure of spectral distortion in TTS and VC, and the word error rate (WER), which reflects the intelligibility of the synthesized or converted speech. For the subjective evaluation, we design listening tests for evaluating the naturalness (NAT) and speaker similarity (SIM) of the synthesized speech. We evaluate the proposed system in two scenarios. In the few-shot TTS scenario, 20 utterances are randomly selected for evaluations. In the few-shot VC scenario, 10 utterances from each source speaker are randomly selected for evaluation. Each utterance is rated by 15 listeners using the mean opinion score (MOS) on a five-point scale. Speech demos are available at https://haitongzhang.github.io/Orchestra/.


 \begin{figure}[t]
    \includegraphics[width=8.5cm, height=5cm]{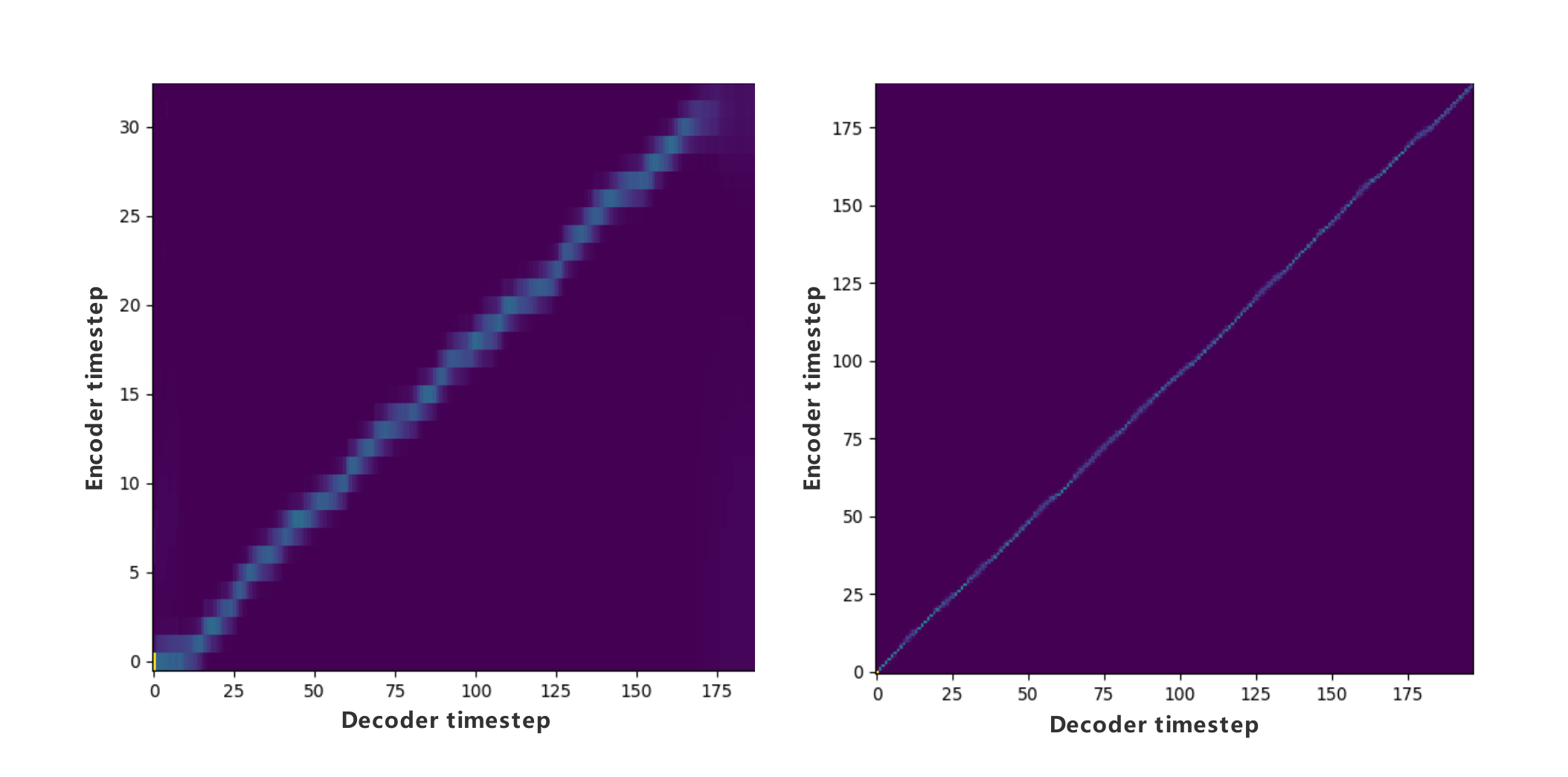}
    \caption{An attention alignment example, the left sub-figure comes from Tacontron2 which uses phonemes as the model's inputs, while the right one comes from MD-Tacotron while using unsupervised linguistic units as inputs.}
    \label{fig:alignment}
 \end{figure}

\begin{figure*}
\centering

\begin{minipage}{0.25\linewidth} \label{mcd}
\centering
\includegraphics[width=1\textwidth]{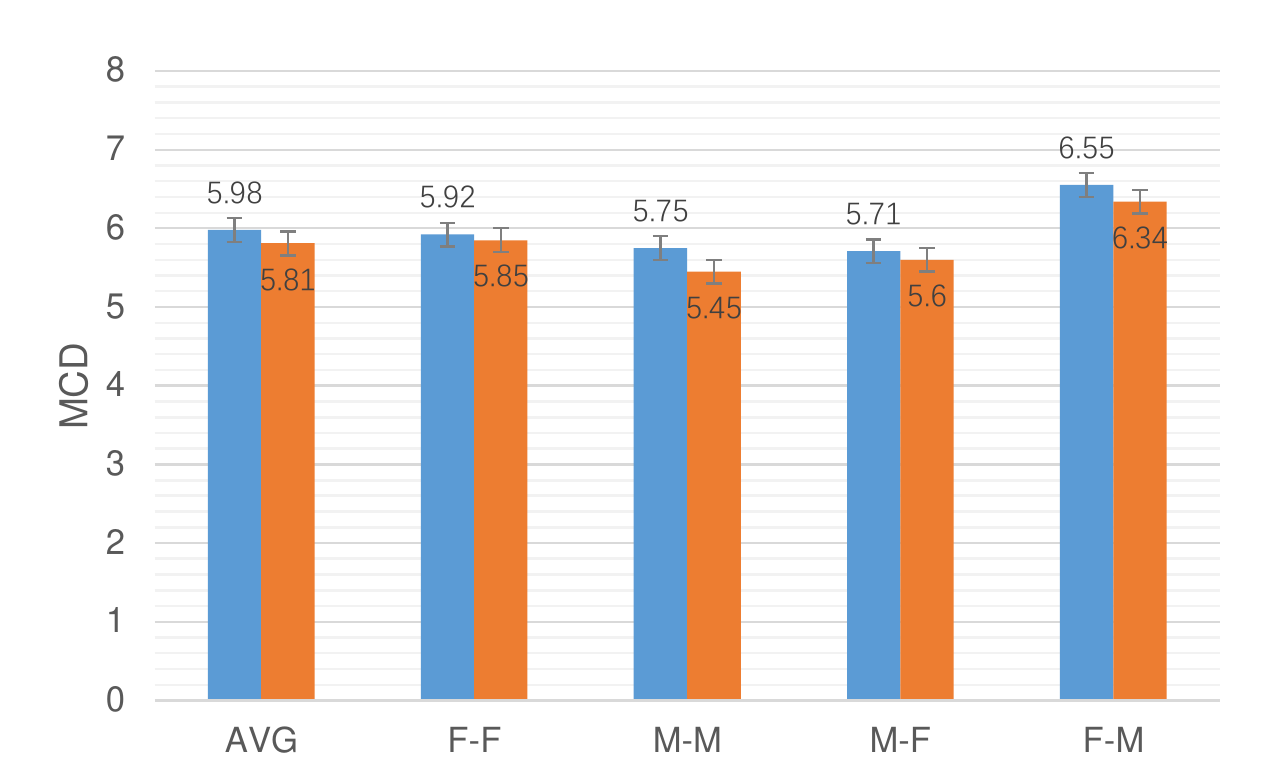}
\subcaption{}
\end{minipage}%
\begin{minipage}{0.25\linewidth}\label{wer}
\centering
\includegraphics[width=1\textwidth]{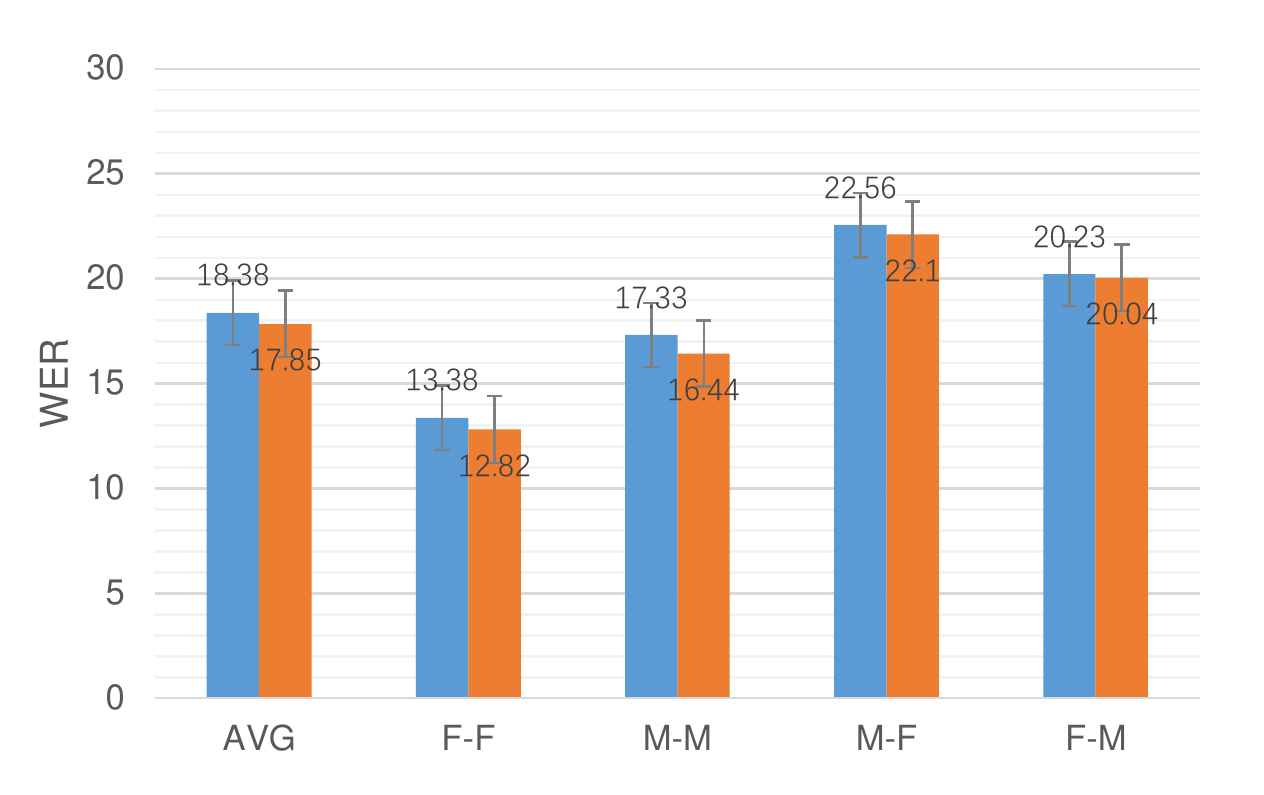}
\subcaption{}
\end{minipage}%
\begin{minipage}{0.25\linewidth}\label{nat}
\centering
\includegraphics[width=1\textwidth]{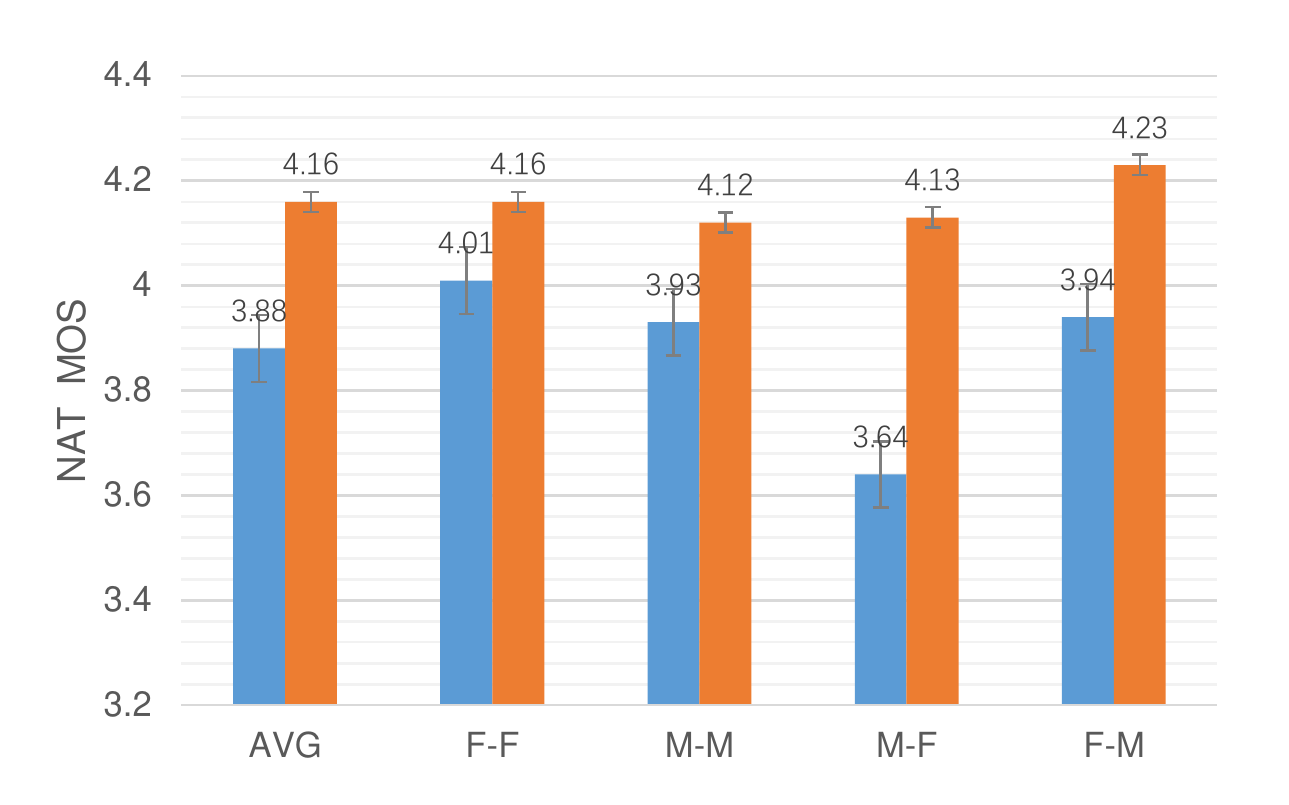}
\subcaption{}
\end{minipage}%
\begin{minipage}{0.25\linewidth}\label{sim}
\centering
\includegraphics[width=1\textwidth]{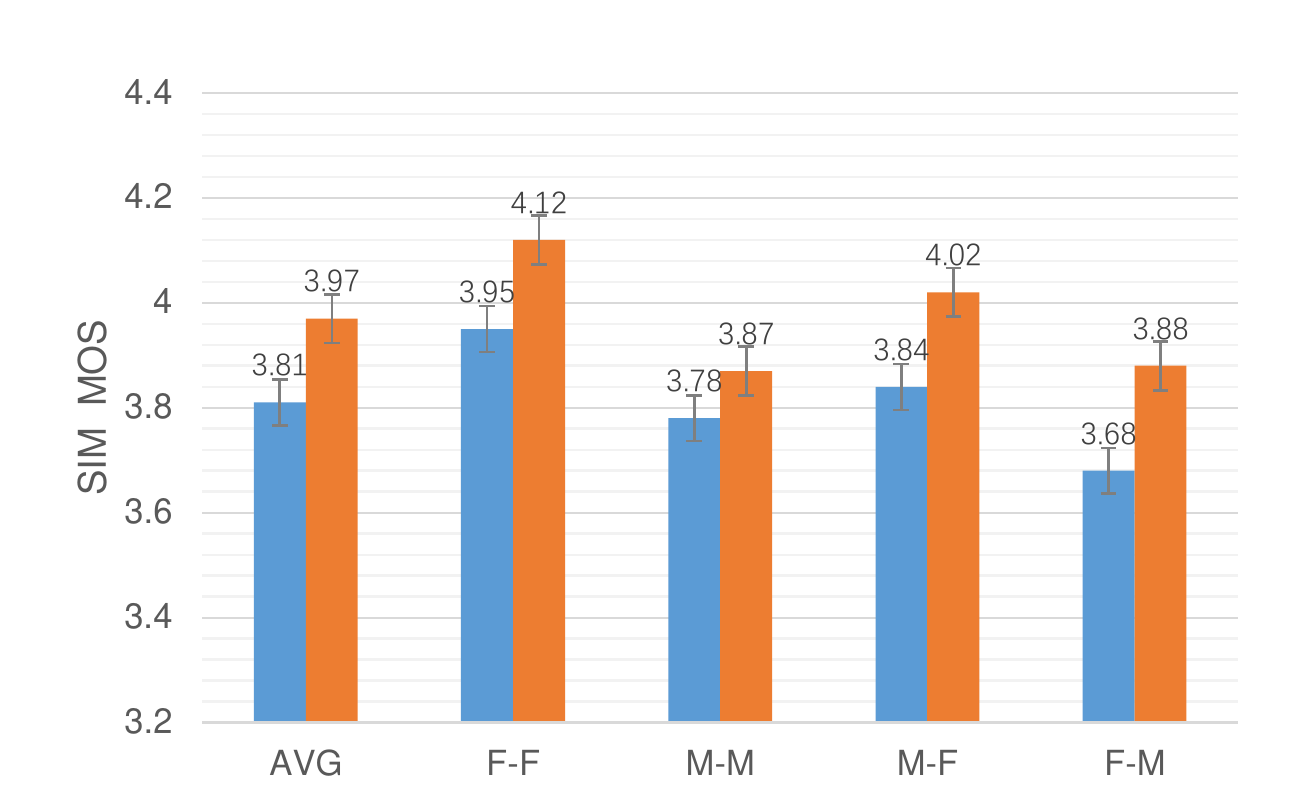}
\subcaption{}
\end{minipage}

\caption{Results of the objective and subjective evaluations for few-shot any-to-one voice conversion; (a) DTW-MCD; (b) WER; (c) Naturalness MOS; (d) Speaker similarity MOS. For DTW-MCD and WER, the lower, the better; for naturalness and speaker similarity MOS, the higher, the better. Blue denotes the VQ-VAE model and orange denotes MD-tacotron.}
\label{res}
\end{figure*}



 

\section{Results and discussion}
\subsection{Few-shot TTS}
We first investigate the effectiveness of our proposed system in the few-shot TTS scenario, where we use speakers p225 and p226 as the target speakers. In this scenario, the evaluation data includes 100 utterances from each target speaker. Since the proposed method takes Tacotron2 as the seq-to-seq module, it is natural to compare it with the Tacotron2 model (the single-modal counterpart) in the few-shot TTS scenario. Thus, the models investigated in this scenario include:

\begin{itemize}
    \item Tacotron:  Tacotron2 pre-trained using $<$ phonemes, audio$>$ paired data from 106 speakers (excluding p225 and p226), then fine-tuned using 2-minute $<$ phonemes, audio$>$ paired data from the target speakers (namely p225 and p226);
    \item MD-Tacotron: The proposed system pre-trained using $<$ phonemes, audio$>$ and $<$ ULUs, audio$>$ paired data from 106 speakers (excluding p225 and p226), then fine-tuned using 2-minute $<$ phonemes, audio$>$ and $<$ ULUs, audio$>$ paired data from the target speakers (namely p225 and p226);
\end{itemize}











To prevent overfitting, we fixed the encoder when fine-tuning the models  \cite{chen2018sample}. The results are given in Table \ref{table:SA}. As shown in Table \ref{table:SA}, the proposed system out-performs Tacotron2 in terms of MCD, WER, and naturalness and achieves a comparable performance to Tacotron2 on speaker similarity.

We attribute the results to two possible reasons. (1) Since ULUs are more fine-grained than phonemes (see Figure \ref{fig:alignment}), the attention mechanism could possibly learn more robust alignment when it takes both fine-grained ULUs and coarse phoneme sequence as inputs. This speculation is confirmed by MD-Tacotron producing less typical alignment errors (i.e., skipping and repeating) than tacotron2 since MD-Tacotron achieves a smaller number of insertion and deletion (while computing the WER) than Tacotron2. (2) We can regard using the unsupervised linguistic units as a way of data augmentation on the textual representation side, which leads to a performance improvement when a limited amount of data is available.


\subsection{Few-shot voice conversion}

In this scenario, we study the performance of the proposed method on the few-shot VC task. We set p225 and p226 as the target speakers and p227 and p228 as the source speakers. Since the proposed method takes VQ-VAE as the extractor of unsupervised linguistic units, it is natural to compare with VQ-VAE Model (the single-modal counterpart) in the few-shot VC scenario. The models investigated in this scenario include:

\begin{itemize}
    \item VQ-VAE:  The single-modal VQ-VAE model pre-trained with speech data from 104 speakers (excluding the source and target speakers), then fine-tuned using 2-minute speech data from the target speakers.
    \item MD-Tacotron: The proposed multi-modal system pre-trained with speech data from 104 speakers (excluding the source and target speakers); then only the seq-to-seq module fine-tuned using 2-minute $<$ ULUs, audio$>$ paired data from the target speakers; 
\end{itemize}

The results are provided in Fig.~\ref{res}. From the objective evaluations (see Fig.~\ref{res} (a) and (b)), we found that the proposed multi-modal system outperforms the single-modal VQ-VAE VC model in MCD and WER. While computing the WER of synthesized speech, we found that the number of substitutions of the proposed system is smaller than that of the VQ-VAE model, which indicates that MD-Tacotron generates more intelligible speech than the VQ-VAE model.

From the subjective evaluations  (see Fig.~\ref{res} (c) and (d)), we found that the proposed system significantly outperforms the baseline VQ-VAE model in naturalness and similarity, which indicates multi-modal learning can improve the few-shot VC performance by a large margin. In addition, we found that the proposed system provides a stable performance on both inter-gender and intra-gender conversion, while VQ-VAE's inter-gender performance is significantly worse than the intra-gender performance.

\section{Conclusion}
This paper proposes a multi-modal voice cloning system by extending Tacotron2 with an unsupervised speech representation module. We verify the effectiveness of the proposed system by comparing it with its single-modal counterparts in both few-shot TTS and VC scenarios. Experimental results reveal that (1) in few-shot TTS, the proposed multi-modal system out-performs its single-modal counterpart (Tacotron2) on MCD, WER, and naturalness, and achieves a comparable performance to Tacotron2 on speaker similarity; (2) in few-shot VC, the proposed multi-modal system out-performs its single-modal counterpart (VQ-VAE) on all evaluation metrics.





\bibliographystyle{IEEEtran}
\bibliography{dataEfficiency}

\end{document}